\documentclass[a4paper,11pt]{article}
\usepackage{aaskaiid}
\usepackage{comment, isomath}
\usepackage{orcidlink}

\def\dn{\Delta \mathbfit{n}}

\newcommand{\hi}{\textrm{H\textsc{i}}}
\newcommand{\secref}[1]{\hyperref[#1]{Section~\ref*{#1}}}
\newcommand{\appref}[1]{\hyperref[#1]{Appendix~\ref*{#1}}}

\def\V{{\mathcal V}}

\newcommand{\uv}[1]{\hat{\mathbfit{#1}}}

\let\svthefootnote\thefootnote
\newcommand\freefootnote[1]{%
  \let\thefootnote\relax%
  \footnotetext{#1}%
  \let\thefootnote\svthefootnote%
}

\title{Interferometric \hi\ Intensity Mapping of the Late Time Universe with SKA-Mid}
\ShortTitle{Interferometric IM}

\author[1]{Aishrila Mazumder$^*$ \orcidlink{0000-0003-3461-496X} }
\author[2]{Zhaoting Chen$^\dagger$ \orcidlink{0000-0002-4965-8239}}
\author[3]{Junaid Townsend \orcidlink{0000-0002-0600-3064}}
\author[3]{Suman Chatterjee \orcidlink{0000-0001-8852-5888}}
\author[4]{Zhixing Li}
\author[1]{Sourabh Paul \orcidlink{0000-0002-8671-2177}}
\author[5]{Reza Ansari \orcidlink{0000-0002-9269-0824}}
\author[1]{Laura Wolz \orcidlink{0000-0003-3334-3037}}
\author[3,6]{Mario G. Santos \orcidlink{0000-0003-3892-3073}}

\affiliation[1]{Jodrell Bank Centre for Astrophysics, Department of Physics and Astronomy, The University of Manchester, Manchester M13 9PL, UK}
\emailAdd{$^*$aishrila.mazumder@manchester.ac.uk}
\affiliation[2]{Institute for Astronomy, The University of Edinburgh, Royal Observatory, Edinburgh EH9 3HJ, UK}
\emailAdd{$^\dagger$zhaoting.chen@roe.ac.uk}
\affiliation[3]{Department of Physics and Astronomy, University of the Western Cape, Robert Sobukwe Road, Bellville, Cape Town 7535, South Africa}
\affiliation[4]{Department of Astronomy, Tsinghua University, Beijing 100084, China}
\affiliation[5] {Université Paris-Saclay, Université Paris Cité, CEA, CNRS, AIM, 91191, Gif-sur-Yvette, France}
\affiliation[6]{South African Radio Astronomy Observatory (SARAO), Liesbeek House, River Park, Gloucester Road, Mowbray, Cape Town, 7700, South Africa}

\abstract{
We discuss the progress towards using the SKA-Mid for interferometric neutral hydrogen (\hi) intensity mapping surveys.
By mapping the distribution of cosmic \hi\ distribution through the 21\,cm line, SKA-Mid will be able to measure the \hi\ power spectrum at small angular separations in interferometric mode.
We review the measurements made from the precursor MeerKAT telescope, using the MeerKAT DEEP2 as well as the MIGHTEE survey data, yielding tentative detection as well as upper limits on \hi\ clustering.
The methdology for MeerKAT can be naturally extended to SKA-Mid.
Forecasts suggest that SKA-Mid AA4 will be able to measure the \hi\ power spectrum with high statistical significance across a wide range of redshifts from $z\sim1.0$ to $z\sim 3.0$, around nonlinear scales $k\sim 1.0\,{\rm Mpc}^{-1}$.
The precise measurements can be used to constrain the properties of \hi\ galaxies, providing a novel window into probing galaxy evolution at $1.0\lesssim z \lesssim 3.0$.

}

\begin{document}
\freefootnote{AM and ZC contribute equally to the chapter.}
\maketitle

\section{Introduction}

In the post-reionisation Universe at $z\lesssim 5.5$ \citep{2022MNRAS.514...55B}, 21\,cm emission line from neutral hydrogen (\hi) can be used to map the three-dimensional distribution of the underlying dark matter. These maps can then be used to constrain the cosmological parameters, particularly the equation of state for dark energy, through galaxy clustering and the Baryon Acoustic Oscillation (BAO) using analysis techniques similar to those employed in optical surveys. \hi\ galaxy surveys with SKA-Mid will provide a large sample of \hi\ selected galaxies up to redshift $z < 0.4$, with precise measurements of redshifts \citep{Nasirudin01.2026.SKA,Mayor01.2026.SKA}. Meanwhile, single dish intensity mapping with SKA-Mid will significantly extend the mappable redshift range for \hi\ to $z \sim 3$ \citep{Cunnington01.2026.SKA}.
  
However, the single dish (SD) \hi\ intensity maps are limited to large angular scales. The angular resolution is limited to the size of the primary beam of the telescope, given by $\sim \lambda/D$, where $D$ is the dish diameter and $\lambda$ is the observed wavelength. For arrays such as the SKA-Mid, with $D=13.5 \mathrm{m}$ and $D=15 \mathrm{m}$ for MeerKAT and SKA-Mid dishes respectively, the primary beam gives an angular resolution of $\gtrsim 1\,$deg at $z=1$. Although this is sufficient to detect the BAO features in the power spectrum up to $z\sim 3$, there are valuable cosmological and astrophysical information at smaller angular scales which will be missed by SD surveys. The SKA-Mid telescope in the interferometric mode will provide information on complementary scales and extending the redshift reach of \hi\ galaxy surveys.

In recent years, significant progress has been made in \hi\ observations with interferometers in the post-reionization Universe. The Canadian Hydrogen Intensity Mapping Experiment (CHIME) has detected large-scale structure signal with \hi\ stacking using eBOSS galaxies and Ly$\alpha$ systems \citep{chime_galaxy, chime_lya}, {and preliminary detection on the auto-power spectrum has been reported \citep{2025arXiv251119620C}}. The upgraded Giant Metrewave Radio Telescope (uGMRT) has placed an upper limit on the post-reionization \hi\ power spectrum \citep{Chakraborty_2021, Elahi01.2026.SKA}. Experiments such as the Hydrogen Intensity and Real-Time Analysis Experiment \citep{2022JATIS...8a1019C} and the Canadian Hydrogen Observatory and Radio-transient Detector \citep{2019clrp.2020...28V} will further explore measurements of \hi\ clustering at BAO scales using interferometry. Interferometry intensity mapping with SKAO has also been studied, in the context of its precursor MeerKAT telescope. Using datasets of various survey areas and depth, measurements of the \hi\ power spectrum have been made, including a tentative measurement \citep{Paul2023} and robust upper limits at $z<0.5$ \citep{2025MNRAS.541..476M}, as well as measurement at $z<0.1$ that matches the expected signal from \hi\ selected galaxies \citep{2026arXiv260223055T}. The precursor measurements are discussed in detail in Section \ref{sec:results}.

The progress in data analysis reveals challenges in mitigating systematics, both in observation and in signal modelling.
In particular, for SKA-Mid, there are significant challenges in dealing with low-level radio frequency interference (RFI) contamination, foreground leakage in high delay modes near the horizon, calibration errors due to field sources around calibrators, and modelling the signal and its covariance at nonlinear scales.
To address these problems, robust simulation tools are needed to understand the observational effects and theoretical framework needs to be established to understand the modelling.

The SKA-Mid will be transformative for \hi\ intensity mapping, providing unprecedented sensitivity and dense coverage of relatively large angular scales.
Interferometric \hi\ intensity mapping benefits from the instrumental power of the SKAO as well as being commensal to continuum surveys.
The existing MeerKAT measurements, for example, utilise datasets that are meant for the MeerKAT DEEP2 observation \citep{2020ApJ...888...61M} and the MeerKAT International GHz Tiered Extragalactic Exploration (MIGHTEE) Survey \citep{mightee}.
It can be foreseen that, without relying on dedicated observation time, SKA-Mid will be able to measure the small-scale \hi\ power spectrum with high statistical significance at a wide range of redshifts $1\lesssim z \lesssim 3$, as we present the forecast in \secref{sec:forecast}.
The precise measurements of \hi\ clustering can be used to probe \hi\ science, shedding light on galaxy evolution, as we discuss in \secref{sec:science}. 
Throughout this chapter, we assume a fixed cosmology corresponding to the model reported in \cite{2020A&A...641A...6P}.

\section{Results from the SKAO precursor MeerKAT telescope}
\label{sec:results}

We describe the results of \hi\ intensity mapping power spectrum measurements using MeerKAT surveys in this section. The MeerKAT array, with its dense core of short baselines and large collecting area, provides excellent sensitivity to large angular scales (low $k_\perp$), which are crucial for 21-cm cosmology. The results so far make measurements at $k$ values $\gtrsim$0.3\,Mpc$^{-1}$, using observations in the L-band, corresponding to \hi\ emission in the redshift range of $0\lesssim z \lesssim 0.5$. 

The studies described here using MeerKAT observations follow a similar power spectrum reconstruction approach. No imaging is performed in this approach, rather visibilities are used directly. They are gridded into $uv$ grids, with $\Delta u$ \& $\Delta v$ values that to make the primary beam width in k$_\perp$ space negligible, thereby minimising the effect of mode mixing. ``Delay-space'' visibilities are obtained by Fourier transforming the gridded visibilities along the frequency axis. The Fourier space visibilities are used to measure the 3D power spectrum. Since the visibilities contain an additive noise bias, it removed either by calculating and removing the expected noise bias or by dividing visibilities into sets of equal but independent time samples, called ``odd'' and ``even'' samples, and correlating them removes the noise bias that is uncorrelated between the samples. These methods give power spectrum measurements via visibility auto-correlation and cross-correlation respectively \footnote{It is to be noted that this is different from auto- and cross-correlations mentioned in context of single dish observations, where they mean self-correlations of intensity maps or between intensity maps and optical galaxy number counts respectively. Instead, here the terms denote either squaring the same visibilities, i.e. $|V_i|^2$ or internal cross-correlation between different sets, $|V_i |*|V_j|$}. The 3D reconstructed power spectrum in Fourier space is given by \citep{2005ApJ...619..678M,2012ApJ...753...81P, 2016ApJ...833..213P}:

\begin{equation}
    P_D(\boldsymbol{k}_{\perp}, k_{\parallel}) = \frac{A_e}{\lambda^2 B} \frac{D^2 \Delta D}{B} \left(\frac{\lambda^2}{2 k_B}\right)^2 \Re \{V_1(\boldsymbol{u}, \tau)V_2^*(\boldsymbol{u}, \tau)\},
\label{eq:vispower}
\end{equation}

where $A_e$ is the effective area of the dish, $\lambda$ is the central wavelength of the corresponding chosen frequency range, $B$ is the total bandwidth, $D$ is the comoving distance calculated at the central redshift, $z$, $\Delta D$ is the comoving width corresponding with the effective bandwidth in frequency, $k_B$ is the Boltzmann constant, and $V_1(\boldsymbol{u}, \tau) = V_2(\boldsymbol{u}, \tau)$ for auto-correlation. All interferometric intensity mapping measurements made with MeerKAT thus far have used the foreground avoidance technique, using only modes above the horizon limit is given by \cite[see, for example][]{2015ApJ...804...14T}:

\begin{equation} \label{eq:horizon_lim}
    k_{\parallel} = \frac{D(z) E(z) H_0}{c(1+z)} \sin(\theta) \ k_{\perp},
\end{equation}

where $E(z) = (H(z) / H_0)^2$, $H_0$ is the Hubble constant, and $\theta$ is the field-of-view (FoV) of the observation. The 1D power spectrum is estimated using an inverse noise variance weighting, taken in logarithmic $k$ bins, $\Delta k_i$, which can be expressed as:

\begin{equation} 
\label{eq:pk_1d_estimator}
    \hat{P}^{i}_D(k) = \sum_j P_D^j w_j  \big /  \sum_j w_j,
\end{equation}

where $w_j = 1/\sigma_{j, \text{TN}}$ is the inverse thermal noise variance weighting. The error on the power spectrum is then estimated using:

\begin{equation} \label{eq:pk_1d_error}
    (\Delta P_D^i)^2 = \sum_j (P^j_D - \hat{P}^i_D)^2 w_j^2 \big / (\sum_j w_j)^2.
\end{equation}

To obtain the weights, $w_j$, multiple thermal noise realisations are simulated, occupying the visibility grid using random values generated from a complex circular Gaussian distribution with a mean of zero, $\mu_{\text{TN}} =0$, and a standard deviation derived from the radiometer equation \citep{2005ApJ...619..678M}:

\begin{equation}
    \sigma_{\text{TN}} = \frac{2k_{\text{B}}T_{\text{sys}}}{A_e \sqrt{\delta\nu \delta t}},
\label{eq:radiometer}
\end{equation} 
where $k_{\rm B}$ is the Boltzmann constant, $T_{\rm sys}$ is the system temperature, $A_{\rm eff}$ is the effective collecting area for one dish, $\delta f$ is the frequency resolution and $\delta t$ is the time resolution.

The standard deviation of the power spectra of these realisations corresponds to $\sigma_{P_{\text{TN}}}$ used in $w = 1 / \sigma^2_{P_{\text{TN}}}$. 
{We note that, the thermal noise simulation is only used to obtain the weights, and the error bar is estimated using the variance of the measured 3D power as described in \autoref{eq:pk_1d_error}.}

Using the aforementioned formalism, the visibility data taken from the tracking observations of the MeerKAT telescope can be used to measure the \hi\ power spectrum, as we review for the rest of this section.

\subsection{Detection of the \hi\ power spectrum from the DEEP2 field}
\label{sp}

\citet{Paul2023} presents the first statistically significant interferometric detection of the \hi\ intensity mapping power spectrum at low redshifts using data from the MeerKAT radio telescope in the L-band. This measurement was performed in the DEEP2 field, selected for its lack of bright continuum sources and long integration time, making it ideal for deep intensity mapping studies.
The analysis used approximately 96 hours of MeerKAT observations collected across nine nights during the telescope's commissioning phase in 2018 \citep{2020ApJ...888...61M}. This data has been used for investigating a wide variety of science cases, for example, sub-mJy source counts \citep{2021ApJ...909..193M}, measuring the local luminosity function of radio sources \citep{2021ApJ...914..126M}, indicating the data quality and reliability. 

The data span two RFI-free frequency windows centred at 1077.5\,MHz ($z \approx 0.32$) and 986\,MHz ($z \approx 0.44$), each with a bandwidth of $\sim$46\,MHz. Our calibration and power spectrum pipeline was specifically developed to enable robust delay-spectrum estimation directly from visibilities, avoiding the need for image cube construction. Even after rigorous RFI flagging, there was evidence of low-level RFI  present in the data. Hence, using a combination of automated CASA-based routines and novel per-baseline delay filtering, we applied two complementary mitigation techniques:
\begin{itemize}
    \item {Baseline-Level Flagging (BLF)}: identifies and removes antenna pairs exhibiting anomalous delay-space power outside the foreground wedge;
    \item {uv-delay Flagging (UVDF)}: excludes individual contaminated modes in the gridded delay-space power spectrum using a 5$\sigma$ threshold over thermal noise.
\end{itemize}

The BLF method identifies and flags contaminated antenna pairs early, on a per-scan basis, before any power-spectrum calculation. By contrast, UVDF applies a threshold after gridding the visibilities, comparing the auto–power spectrum to the expected thermal noise. As a result, BLF, being closer to traditional radio-data flagging, tends to be more aggressive, while UVDF, applied nearer to the final power-spectrum stage, typically flags less. Both approaches yield consistent results.

\begin{figure}
    \centering
	\includegraphics[width=0.8\columnwidth]{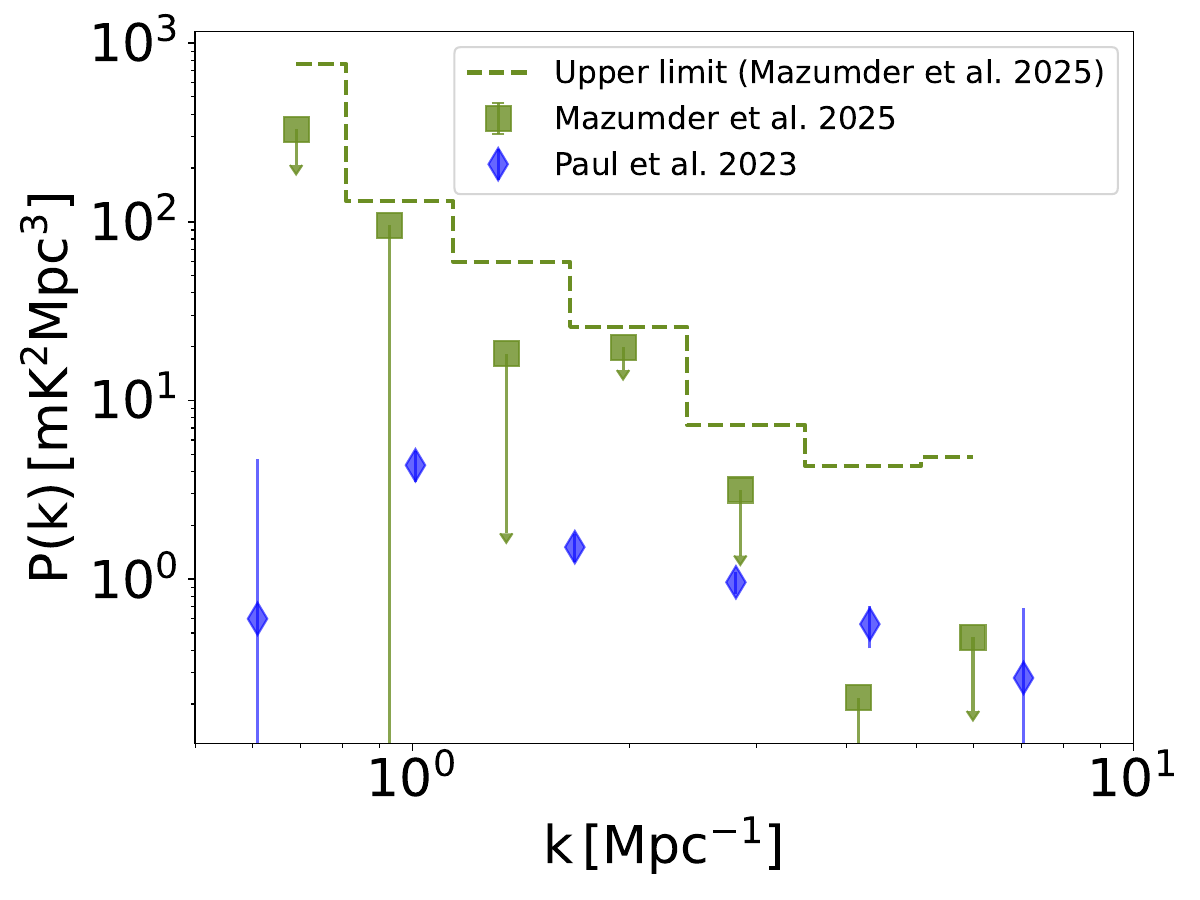}
    \caption{\hi\ power spectrum estimates from MeerKAT observations at z$\sim$0.44. The purple squares represent the upper limits from \hi\ power spectrum obtained using cross correlation, and purple dashed lines represent the absolute upper limits using MIGHTEE data \citep{2025MNRAS.541..476M}. Blue diamonds show the tentative detection using the MeerKAT DEEP2 data reported in \cite{Paul2023}.}
    \label{fig:meerkat}
\end{figure}

We employed a conservative foreground avoidance strategy, applying a cut at $k_\parallel > 0.3k_\perp$ to ensure robustness against residual contamination. Following these mitigations, we obtained clear detections of the 3-d \hi\ auto-power spectrum:
\begin{itemize}
    \item $\sim$5.9$\sigma$ significance at $z \approx 0.32$, and
    \item $\sim$9.2$\sigma$ significance at $z \approx 0.44$.
\end{itemize}

\autoref{fig:meerkat} shows the 1D \hi\ power spectrum at $z \approx 0.44$ in blue diamonds, overplotted with MIGHTEE results (described in Section \ref{mightee} below). These measurements represent the first detection of interferometric auto-power spectra at these redshifts, achieved without relying on cross-correlation with galaxy surveys. The amplitude and scale dependence of the measured power spectra are consistent with predictions from halo-based models, and the signal is robust under multiple calibration and systematic error tests.

\subsection{Upper limits of the \hi\ power spectrum with multiple fields from the MIGHTEE survey}
\label{mightee}

\cite{2025MNRAS.541..476M} presents a follow-up to the DEEP2 power spectrum measurement using the MeerKAT International GigaHertz Tiered Extragalactic Exploration, i.e. the MIGHTEE survey \citep{mightee}. It covers $\sim$20 square degrees with a total of 1000 hours of observation time distributed over four well-known extragalactic deep fields.  A subset of the data covering $\sim$100 hours in the COSMOS field was used for this study \citep{2024MNRAS.534...76H}. The observations span $\sim$4 square degrees, consisting of 15 pointings with 94.2 hours on-source. Owing to the different phase centres of each pointing, they sample slightly different parts of the sky. Hence, the gridded, delay-transformed visibilities were incoherently averaged in the power spectrum domain for the final measurements. 

The frequency range chosen in this work covers the same RFI free window centred at $\sim$986\,MHz (or \hi\ emission at $z\approx0.44$) as that in \cite{Paul2023}. Since incoherent averaging is the so-called ``square-then-average'' method, contrary to coherent averaging or the ``average-then-square'' method, the noise scales down as $1/\sqrt{N}$, since N independent measurements (pointings) are averaged after squaring, as opposed to averaging N independent measurements (visibilities) and then squaring, which reduces the noise by 1/N. Hence, we get a root-mean-squared thermal noise fluctuation $\sim$10\,mK using an observation spread over 15 pointings (for a single deep observation of the same time, it would be $\sim$5\,mK).

The calibrated visibilities for each observing night were gridded in the $uv$-plane with a grid size $60\lambda$. The gridded cubes were Fourier transformed along the frequency axis to produce a ($u,v,\tau$) cube, where  $\tau$ is the Fourier conjugate of frequency, or the delay. The 3D delay power per pointing $P_D$ given by \autoref{eq:vispower} is weighted by the respective baseline density for each pointing, to account for the differences in baseline coverage among the pointings. This gives the final averaged 3D power:

\begin{equation}
    P_{D} (\textbf{\textit{k}}_{\perp}, k_{\parallel}) =\frac{\sum\limits_{i=1}^{15} w_{i}(\textbf{\textit{k}}_{\perp}, k_{\parallel})\,P_D^i(\textbf{\textit{k}}_{\perp}, k_{\parallel})}{\sum\limits_{i=1}^{15} w_{i}(\textbf{\textit{k}}_{\perp}, k_{\parallel})} ,
    \label{ic_eq}
\end{equation}

where the $i$ is over all 15 pointings and $P_D^i(\textbf{\textit{k}}_{\perp}, k_{\parallel})$ is the delay power for the $i$th pointing, weighted by the factor $w_{i}$ which is the baseline density per 3D $k$ pixel. Finally, the incoherently averaged 3D power spectrum is binned into spherical and cylindrical bins. The data usability and feasibility of using incoherent averaging were checked through jackknife tests, checks for power distribution per 1D k-bin and varying the foreground avoidance window. These tests showed that while residual systematics may be present in some pointings, on combination and cross-correlation, most systematic contributions reduce considerably. 

Employing the same foreground avoidance window described in Section \ref{sp}, we set the first upper limits on the \hi\ power spectrum. It should be noted that, unlike \cite{Paul2023}, we do not use any baseline level or uv-delay space flagging in this work. Using both auto-correlation and cross-correlation visibilities (see \autoref{fig:meerkat}), we find consistent upper limits of 29.8 \,mK$^{2}$Mpc${^3}$ from the 2$\sigma$ cross-correlation measurements and 25.82\,mK$^{2}$Mpc${^3}$ from auto-correlation at $k\sim$2\,Mpc$^{-1}$. The data used here constitutes a small fraction of the MIGHTEE survey and demonstrates that combined analysis of the full MIGHTEE survey can potentially detect the \hi\ power spectrum at $z\lesssim0.5$, in the range $0.1\,\textrm{Mpc}^{-1} \lesssim k \lesssim 10\,\textrm{Mpc}^{-1}$ or quasi-linear scales.

\autoref{fig:meerkat} shows the \hi\ clustering power spectrum upper limits from auto- (dashed purple lines) and the cross- (purple squares) correlation of MIGHTEE-COSMOS observations.
{The MIGHTEE COSMOS data presented in this subsection has a similar amount of observational time with the DEEP2 field. However, the observation consists of 15 pointings whereas the DEEP2 data is one single pointing, and the final measurement is an incoherent averaging of the power spectrum measurements from each pointing. Therefore, the signal-to-noise ratio of the MIGHTEE measurements is lower, with only upper limits reported. Nevertheless,}
these consistent results demonstrate the viability of MeerKAT as a precision interferometric instrument for cosmological \hi\ intensity mapping {across multiple pointings over a large survey area}, providing a strong foundation for future observations with the SKA-Mid array.

\subsection{A comparative analysis of \hi\ clustering between intensity maps and direct detections}
\label{subsec:cosmos_lowz_hi_im}

MeerKAT observations have also been used to test the robustness of using the interferometric \hi\ intensity mapping in the low-redshift Universe ($z<0.1$). In this redshift regime, large-scale \hi\ galaxy surveys are feasible, since the sensitivity requirements are less strenuous on radio telescopes. As one goes to higher redshifts, this becomes harder to do, making intensity mapping ideal for probing the \hi\ content in the post-EoR Universe beyond $z=0.1$.

With the MIGHTEE Early-Science COSMOS field data \citep{2016mks..confE...6J}, there are known \hi\ galaxy detections. \cite{2026arXiv260223055T} explores an interesting way to test the performance of intensity mapping, using known \hi\ sources. Here, both interferometric observations and \hi\ selected galaxies in the same field of view are used to estimate and compare \hi\ power spectra from each of these data sets. Such tests can be used for (a) testing how much neutral gas each of the methods (intensity mapping and galaxy detections) misses, and (b) testing the robustness of the methodology developed for processing visibility data for \hi\ power spectrum estimation (used in \citealt{Paul2023, 2025MNRAS.541..476M}). Further, such analyses allow us to probe the \hi\ clustering present in both the galaxy samples and the interferometric data sets on ultra-small scales ($3 \lesssim k \lesssim 40$ Mpc$^{-1}$).

\begin{figure}
    \centering
    \includegraphics[width=\columnwidth]{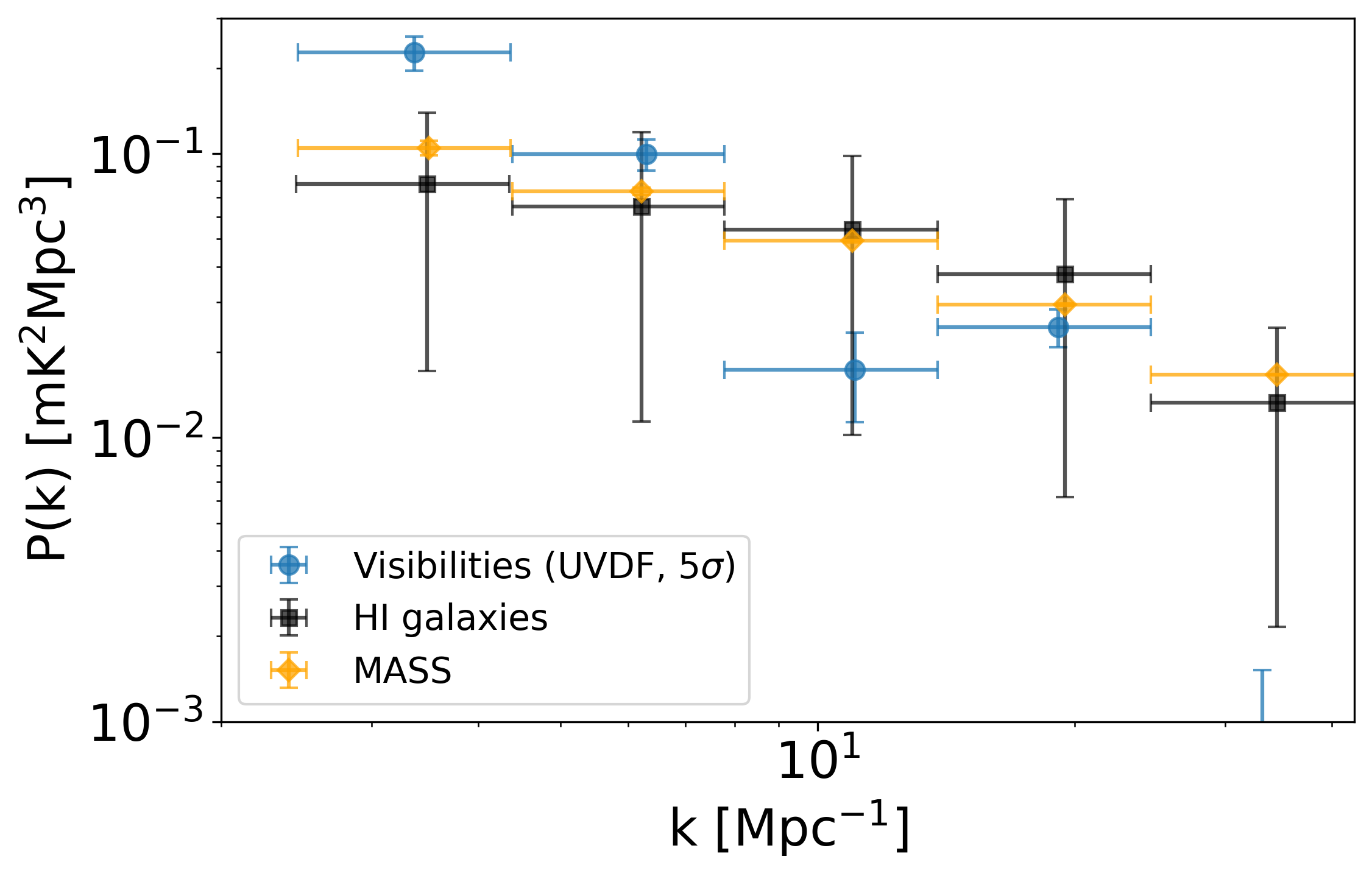}
    \caption{{\hi\ power spectrum estimates from MeerKAT observations at $z \lesssim 0.1$ using the MIGHTEE Early-Science COSMOS data \citep{2026arXiv260223055T}.
    The blue circles denote the measurement from visibility data, with a 5$\sigma$ UVDF cut.
    The black squares denote the measurement from the \hi\ galaxy catalogue. Orange diamonds denote measurements from simulated visibilities (MASS) using the \hi\ galaxy catalogue.}
    }
    \label{fig:pk_results}
\end{figure}

In this analysis, the data used consists of 3 nights on the same central region of the COSMOS field, combining observations from both Early-Science data \citep{2022MNRAS.509.2150H} and MIGHTEE spectral line data \citep{2024MNRAS.534...76H}, with a total observation time of $\sim$17.45\,hours and a redshift range of 0.02 $< z <$ 0.066. Conversely, \citep{2025MNRAS.541..476M} use all 15 pointings on the COSMOS field described in \citep{2024MNRAS.534...76H}. The calibrated visibility data were used to perform the cross-correlation, as discussed above. The UVDF approach described in Section \ref{sp} is adopted, flagging delay modes with a 5$\sigma$ threshold. The results are shown in \autoref{fig:pk_results} as the blue circle points for the 5$\sigma$ cut.

Since this analysis aims txo use the \hi\ galaxy detections of the same data to compare and validate the power spectrum measurements, we also require the \hi\ selected galaxy catalogue. This catalogue is generated following \cite{2021A&A...646A..35M}, with a measurement of the \hi\ mass and the velocity width of the \hi\ profile for each galaxy. We then use two methods to estimate the \hi\ power spectrum from the galaxy catalogue. First, we grid the measured \hi\ emission into a rectangular grid, and perform power spectrum estimation directly on this grid. The results are shown in \autoref{fig:pk_results} as the black square points (``HI galaxies''). Second, we use the source information to simulate \hi-only visibilities using the MeerKAT All-Sky Simulator (MASS; see Section 5.1 of \citealt{2026arXiv260223055T}). The \hi-only visibilities are then used to perform the power spectrum estimation and compare to the results from the data, as shown in the orange diamond points (``MASS'') in \autoref{fig:pk_results}.

As one can see, for the presented range of $k$ scales, the measurements of the power spectrum amplitude from the intensity maps agree with the overlapping \hi-selected galaxy power spectrum. This analysis indicates the robustness of the approach adopted in the MeerKAT data analysis for interferometric intensity mapping.

In conclusion, interferometric intensity mapping with MeerKAT has successfully demonstrated that both deep single-pointing and multi-pointing wide-area surveys are capable of measuring the clustering signal from \hi. The precursor MeerKAT data analysis indicates that interferometric intensity mapping with SKA-Mid holds great potential for measuring the \hi\ power spectrum at small scales across a wide range of redshifts.
It is worth noting that these measurements are made using data designed for continuum and emission line galaxy science. In principle, any tracking observations with SKA-Mid can be used to perform power spectrum analysis, yielding great synergistic potential with other science goals of the SKAO.

\subsection{MeerKAT All-Sky Simulator (MASS)}
\label{subsection:mass_sims}

Besides observations, there have also been developments in simulating and modelling systematics in interferometric data. While several such examples are present for EoR, one example in the post-reionization regime is the MeerKAT All-Sky Simulator (MASS).

MASS is designed to simulate interferometric visibilities for a given astrophysical sky model. This package is developed for large-scale \hi\ intensity mapping with interferometric observations using SKA-Mid precursor MeerKAT.  MASS can be used for simulation based forward modelling approaches where realistic mock data is generated to study the impact of systematics and other complex processes applied to the observed data. In its current format, the MASS sky model can be a point-like source catalogues (Dirac delta functions) or a \texttt{HEALPix} \citep[Hierarchical Equal Area isoLatitude Pixelization of a sphere;][]{Gorski2005} map. MASS can also incorporate analytical models as input and construct source catalogues or \texttt{HEALPix} maps. With a given antenna configuration or baseline distribution, the package generates measurement sets that can be directly used in the data processing pipeline.


The visibilities are simulated using:

\begin{equation}
\V(\mathbfit{\theta}, \mathbfit{U},\nu) = Q_{\nu}\,  \sum _{q=0}^{N - 1} \,\Delta \Omega_{\uv{n}_q} \, T(\uv{n}_q,\nu)
\, A(\dn_q,\nu)  e^{2 \pi i \mathbfit{U} \cdot \dn_q},
  \label{eq:v8}
\end{equation}

where the antenna primary beam (PB) pattern $A\left(\dn_q,\nu \right)$ quantifies how the individual antenna responds to the signal from different directions $\hat{\mathbfit{n}}$ on the sky, $\hat{\mathbfit{m}}$ refers to the direction in which the antennas are pointing with $\Delta \hat{\mathbfit{n}}_q =\hat{\mathbfit{n}}_q -\hat{\mathbfit{m}}$, $Q_{\nu}= 2 k_B / \lambda^2$ is the conversion factor from brightness temperature to specific intensity in the Rayleigh - Jeans limit, $T\left(\hat{\mathbfit{n}}_q,\nu \right)$ is the brightness temperature distribution on the sky and  $\Delta \Omega_{\uv{n}_q}$ is the elemental solid angle in the direction $\uv{n}_q$. The summation can be over \texttt{HEALPix} pixel or the source catalogue. The primary advantage of this simplistic approach is that one can include different components at scales that they dominate. For instance, it is seen that bright point sources far away from the central lobe of the telescope PB will impact the \hi\ signal that is in the primary lobe of the beam. Current simulation capabilities of MASS include diffused galactic synchrotron emission, free-free emission, extra-galactic point sources, the Sun, Radio frequency interference (RFI, narrow and broad band), noise, gain error, and primary beam with side-lobes that extend to the horizon. However, the current version of MASS is not suitable for simulating small scale diffused emission.

\section{Forecasts of power spectrum measurements for SKA-Mid}
\label{sec:forecast}

This section describes the forecasts from SKA-Mid for interferometric intensity mapping.  \hi\ intensity mapping experiments require good coverage in spatial frequencies, i.e. the uv plane to ensure sufficient sensitivity to the faint cosmological signal, and to limit the impact of mode mixing. Hence, most of the dedicated interferometers for intensity mapping are configured as packed arrays.  Although SKA-Mid is a general-purpose interferometer designed to reach high angular resolution, the configuration of the central part of the array, with baseline lengths $\lesssim \mathrm{1\,km}$, is suitable for intensity mapping and provides good coverage of the uv plane.
As discussed in \secref{sec:results}, intensity mapping can be performed using continuum survey data of SKA-Mid, with multiple pointings of relatively deep integration time.
Here, a demonstration is given of how these features translate to signal-to-noise expectations and the detectability of the \hi\ clustering power spectrum at redshifts beyond 0.5.

\subsection{Signal-to-Noise Estimates}
\begin{figure}[h]
    \centering
	\includegraphics[width=\columnwidth]{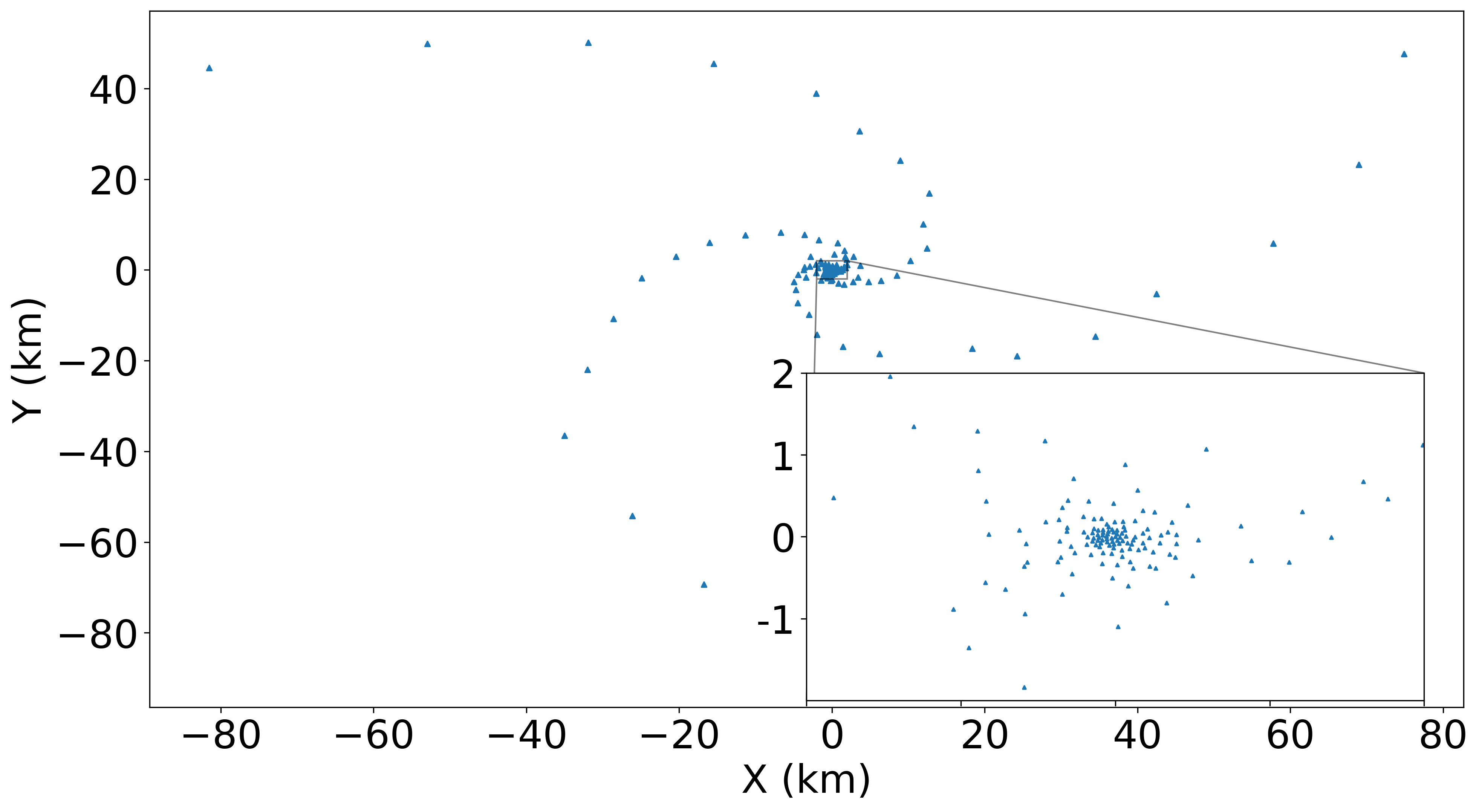}
    \caption{Left: $(u,v)$ plane values of all dish pairs for the SKA-Mid array with 197 dishes at $700 \mathrm{MHz}$. Right: $(u,v)$ values for baselines with $\sqrt{u^2+v^2} < 1500$}
    \label{fig:config}
\end{figure}

Using the AA4 array configuration specified in \cite{2019arXiv191212699B}\footnote{\url{https://gitlab.com/ska-telescope/ost/ska-ost-array-config}}, \autoref{fig:config} shows the array layout for the 197 SKA-Mid dishes, with a zoom-in on the dense core within 2\,km of the array centre. It shows that 118 out of 197 dishes are within the 1\,km of the array centre, corresponding to $\sim 36\%$ of the baselines.  Thus, it will be possible to map out the \hi\ distribution with $\sim  1 \mathrm{arcmin} $ resolution at redshift $z=1$ with high sensitivity using the SKA-Mid array.

\begin{figure}[h]
    \centering
	\includegraphics[width=0.45\columnwidth]{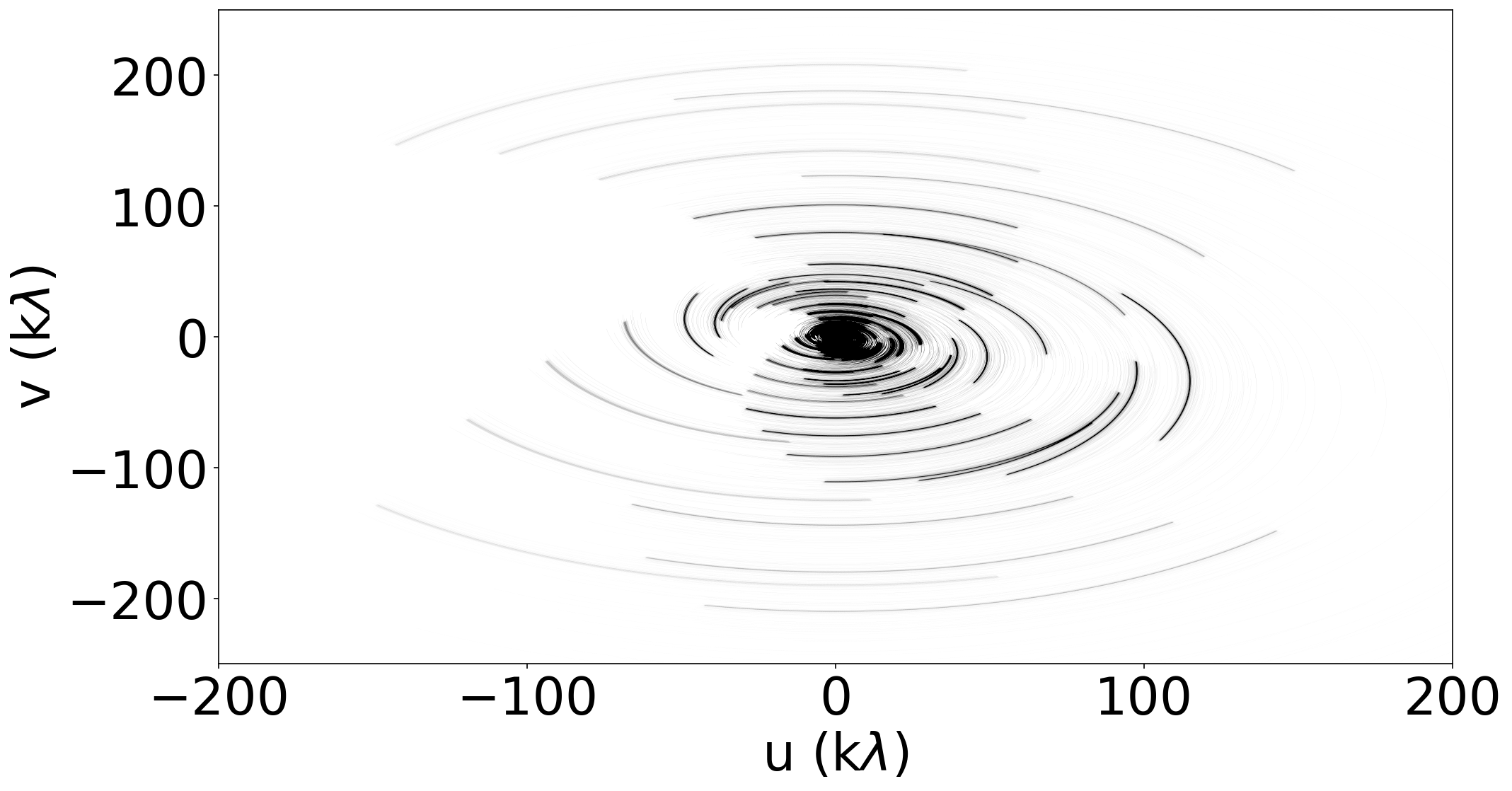}
	\includegraphics[width=0.45\columnwidth]{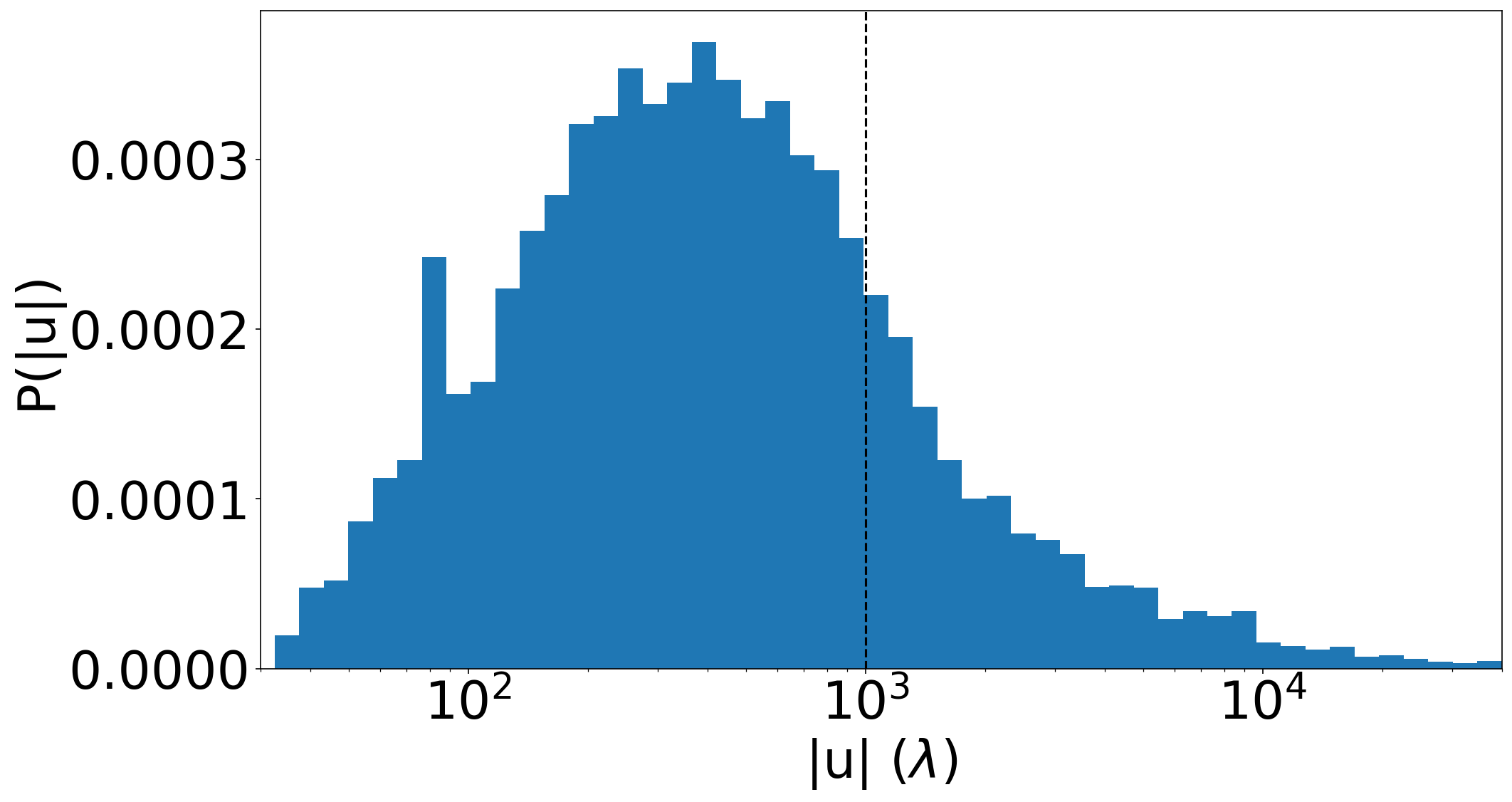}
    \caption{Left: The redshift distribution of the baselines for SKA-Mid AA4 for a tracking observation.  Right: The probability distribution of the uv distance $|u| = \sqrt{u^2+v^2}$ for SKA-Mid AA4. The dashed line denotes $|u|=1000\,\lambda$, which is a crude threshold for baselines that probe scales useful for clustering.}
    \label{fig:uv}
\end{figure}

To calculate the expected uv coverage and the corresponding signal-to-noise level for the survey, a 4hr tracking observation pointing at the DEEP2 field is simulated. It is assumed that the visibility data will have a time resolution of 10\,s. The resulting uv coverage is shown in \autoref{fig:uv}.

The radiometer equation, as described in \autoref{eq:radiometer}, is used to simulate thermal noise realisations, and forecasts are presented for three redshift bins at $0.5<z<1.5$ ($z\sim 1$), $1.5<z<2.5$ ($z\sim 2$) and $2.5 <z <3.0$ ($z\sim 3$), with a frequency resolution of $0.38\,$MHz for each bin\footnote{Note that the frequency resolution of the visibility data can be higher, and the choice of frequency resolution in our forecast corresponds to the visibility data being averaged down to a lower resolution.}. The total observation datasets are assumed to consist of 100 trackings of different pointings, each with 10\,hr of observation time.

Then the visibilities are gridded to a uv grid following \cite{Paul2023}. A grid size of $\Delta u = \Delta v = 20\,\lambda$ is used here.
The thermal noise generated per grid follows
\begin{equation}
    \sigma_{\rm TN}^{\rm grid} = \sigma_{\rm TN} \sqrt\frac{{t_{\rm track}}}{{t_{\rm field}}\,{N_{\rm baseline}^{\rm grid}}},
\end{equation}
where $\sigma_{\rm TN}$ is the noise level for a single baseline following \autoref{eq:radiometer}, $t_{\rm track}=4\,$hr is the observation time of the simulated tracking, $t_{\rm field} = 10\,$hr is the total observation time of one field.

Then the thermal noise realisations are generated and the brightness temperature power spectrum is calculated based on \autoref{eq:vispower} for 128 realisations.
The 3D power spectrum is averaged into the 1D power spectrum. The 1D $|k|$-bins are logarithmically spaced between 0.075$\,\rm Mpc^{-1}$ to 10$\,\rm Mpc^{-1}$ for $z\sim 1$, 0.05$\,\rm Mpc^{-1}$ to 8$\,\rm Mpc^{-1}$ for $z\sim 2$ and 0.04$\,\rm Mpc^{-1}$ to 6$\,\rm Mpc^{-1}$ for $z\sim 3$.
Finally, following equation \autoref{eq:horizon_lim}, a horizon limit corresponding to $\theta = 5\,$deg is chosen.
This is a conservative choice as the FoV of the SKA-Mid dishes will be $\sim$1\,deg, which effectively considers potential foreground leakage at high delay.
The 3D $k$-modes above the limit is then used for averaging into 1D.
The variance of the 1D power spectrum among the realisations is used for the thermal noise variance so that
\begin{equation}
    \sigma_{P_{\rm TN}}^2 = {\rm var}\big[P_{\rm TN}^i\big] / N_{\rm field},
\end{equation}
where ${\rm var}\big[P_{\rm TN}^i\big]$ is the variance among the realisations for a single field, and the final thermal noise variance is the result of incoherent averaging of all the fields.

Additionally, the fiducial signal $P_{\rm \hi}$ as well as its variance are both calculated. The chosen fiducial model corresponds to thee halo model described later in \secref{subsec:halomodel}.
The signal variance is then $P_{\rm \hi} / \sqrt{N_k}$, where $N_k$ is the number of $k$-modes in each 1D k-bin.
The correlations between different $k$-bins are ignored for simplicity.

\begin{figure}[h]
    \centering
	\includegraphics[width=0.49\columnwidth]{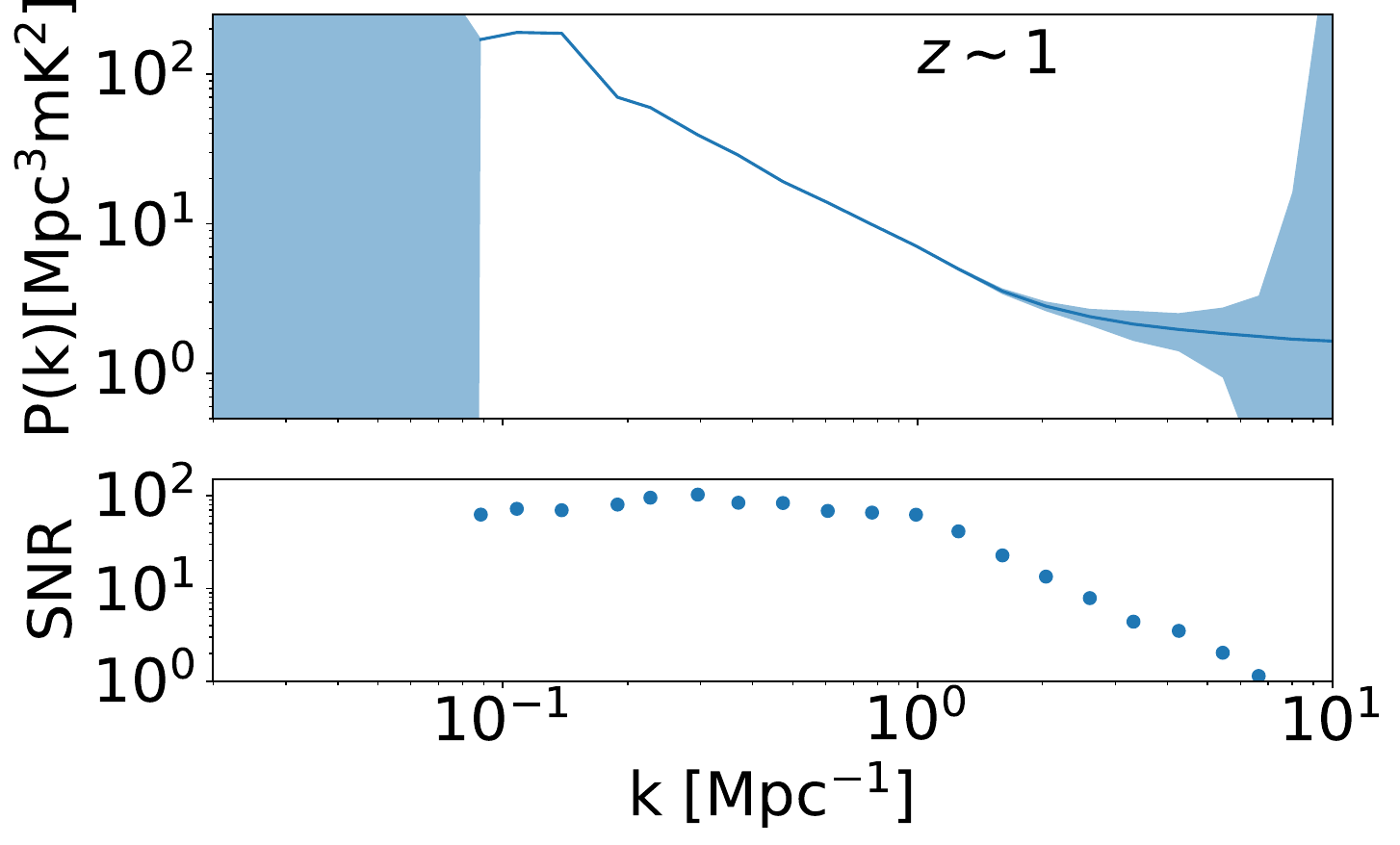}
	\includegraphics[width=0.49\columnwidth]{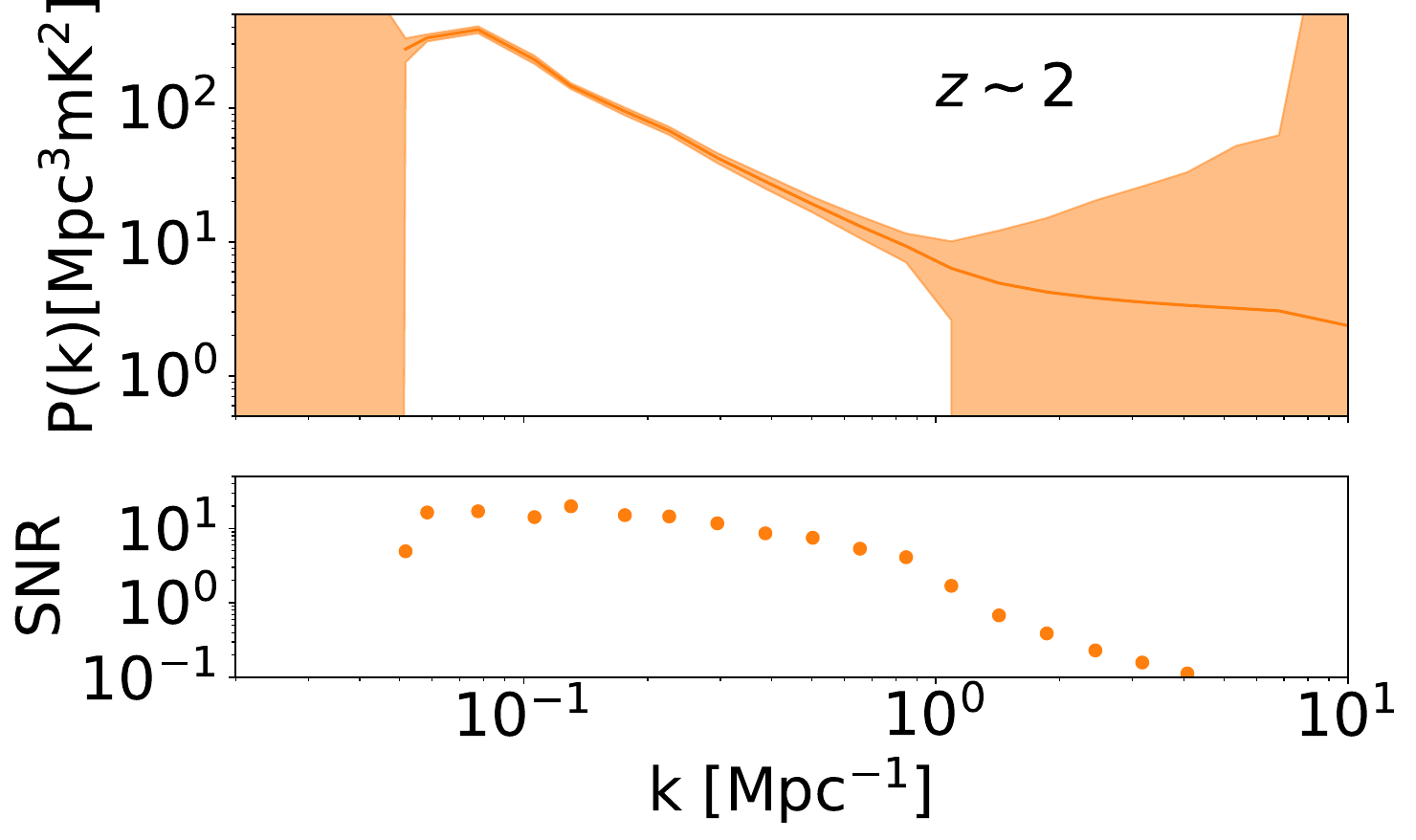}
	\includegraphics[width=0.49\columnwidth]{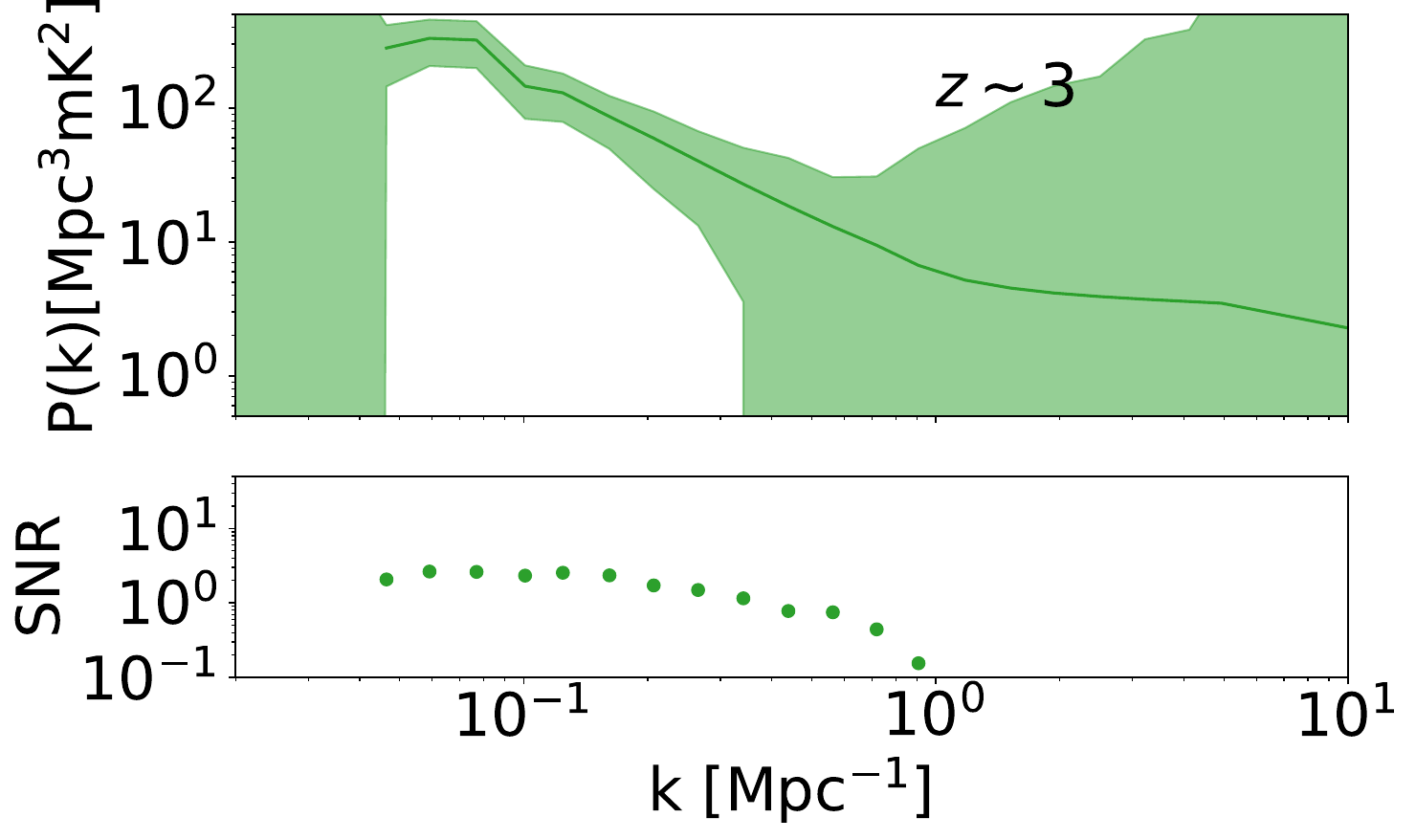}
    \caption{Forecasts for the interferometric \hi\ intensity mapping measurements for SKA-Mid AA4 configuration. The top half of each panel shows the fiducial \hi\ power spectrum and the expected 1$\sigma$ measurement error. The bottom half of each panel shows the expected signal-to-noise ratio of each 1D $k$-bin. The redshift bins shown are $0.5<z<1.5$ ($z\sim1$), $1.5<z<2.5$ ($z\sim2$) and $2.5<z<3.0$ ($z\sim3$).}
    \label{fig:snr}
\end{figure}

The results are shown in \autoref{fig:snr}. For SKA-Mid AA4, the \hi\ clustering with interferometric observations across a wide range of scales can be probed.
High accuracy measurements, with a signal-to-noise ratio larger than 10, can be achieved around $0.1 \lesssim k\lesssim 1\,{\rm Mpc^{-1}}$ for $z\sim 1$ and $z\sim 2$.
For $z\sim 1$, scales up to $k\sim 10 \,{\rm Mpc^{-1}}$ can be probed.
For $z\sim 3$, detection of the \hi\ power spectrum can be made at $k\sim 0.1\,{\rm Mpc^{-1}}$.





\section{Anticipated science outputs from interferometric intensity mapping using SKA-Mid}
\label{sec:science}

In \secref{sec:forecast}, it is demonstrated that SKA-Mid can be used to probe the \hi\ clustering with high accuracy.
The measurements can then be used to probe \hi\ science, and the potential science outputs are discussed in this section.

\subsection{Halo Model}
\label{subsec:halomodel}
The measured \hi\ power spectrum at quasi-linear and non-linear scales can be used to constrain the halo model of cosmic neutral hydrogen \citep{2021MNRAS.502.5259C}. In halo model formalism, the power spectrum is modelled as a combination of a 2-halo, a 1-halo term and a shot noise term \citep{2002PhR...372....1C}.
Following \cite{2021MNRAS.502.5259C}, the \hi\ power spectrum can be modelled as:
\begin{equation}
    P_{\rm \hi}(k_\perp,k_\parallel) = P_{\rm 2h}(k_\perp,k_\parallel) + P_{\rm 1h}(k_\perp,k_\parallel) + P_{\rm SN}(k_\parallel).
\end{equation}
The 2-halo term is
\begin{equation}
\begin{split}
P_{\rm 2h}(k,k_\parallel) =& C_{\rm \hi}^2 \bigg[\int{\rm d}m\,n(m) b(m)\langle M_{\rm \hi}(m) \rangle u_{\rm \hi}(k|m) \bigg]^2(1+\frac{fk_\parallel^2}{b_{\rm \hi}^0k^2})^2 \frac{P_{m}(k)}{{1+(k_\parallel\sigma_p)^2}/2},
\end{split}
\end{equation}
where $m$ integrates over the halo mass range, $C_{\rm \hi}$ is the conversion factor from \hi\ mass density to temperature, $n(m)$ is the halo mass function \citep{Tinker_2008}, $b(m)$ is the linear halo bias \citep{Tinker_2010}, $\langle M_{\rm \hi}(m) \rangle$ is the \hi-halo mass relation, $u_{\rm \hi}(k|m) = \mathcal{F}(\rho_{\rm \hi}(r|m))$ is the normalised Fourier-transform of the density profile $\rho_{\rm \hi}(r|m)$, $k=\sqrt{k_\perp^2+k^2_\parallel}$ and $b_{\rm \hi}^0$ is the linear \hi\ bias. The $(1+({fk_\parallel^2})/({b_{\rm \hi}^0k^2}))^2$ term denotes the Kaiser effect \citep{Kaiser_1987}, and ${1+(k_\parallel\sigma_p)^2}/2$ denotes the finger-of-god (FoG) effect \citep{Jackson_1971}.
$f$ is the growth rate and $\sigma_p =\sigma_v(1+z)/(Hf) $ where $\sigma_v$ is the velocity dispersion and $H$ is the Hubble parameter.
$P_{\rm m}(k)$ is the matter power spectrum. 
 
 The linear \hi\ bias is calculated as
\begin{equation}
    b_{\rm \hi}^0 = \frac{\int{\rm d}m\,n(m) b(m)\langle M_{\rm \hi}(m) \rangle}{\int{\rm d}n\,n(m) \langle M_{\rm \hi}(m) \rangle},
\end{equation}
while the 1-halo term can be modelled as
\begin{equation}
\begin{split}
    P_{\rm 1h}(k,k_\parallel) =C_{\rm \hi}^2 \int {d}m \,n(m)\langle M_{\rm \hi}(m)\rangle ^2 u_{\rm \hi}(k|m)\frac{1}{{1+(k_\parallel\sigma_p)^2}/2} .
\end{split}    
\end{equation}

The shot noise term $P_{\rm SN}(k_\parallel)$ can be written as
\begin{equation}
    P_{\rm SN}(k_\parallel) = \frac{P_{\rm SN}^0 }{1+(k_\parallel\sigma_p)^2/2}.
\end{equation}
The shot noise in the \hi\ power spectrum is also attenuated by the FoG effect. This is because the 21\,cm line is broadened by the internal velocity dispersion of \hi\ galaxies. The broadening breaks the point-source assumption along the line of sight, smearing the overall power spectrum along the $k_\parallel$ direction, including the shot noise (e.g. \citealt{2020ApJ...895...34Z}).

Using halo model, we can also calculate the average \hi\ density as
\begin{equation}
\Omega_{\rm \hi} = \int {\rm d}n\; n(M)\langle M_{\rm \hi}(m) \rangle\big/\rho_c,
\label{eq:omegahi}
\end{equation}
where $\rho_c$ is the critical density of the Universe at $z=0$.

We follow the parameterisation in \cite{Paul2023} so that
\begin{equation}
\begin{split}
        \langle M_{\rm \hi}(m)\rangle = A_{\rm \hi}\big(\frac{m}{M_0}\big)^\beta,\,\\ \rho_{\rm \hi}(r|m) = \rho_0 \big(\frac{r}{r_0}\big)^\gamma {\rm exp}[-ar/r_0],
\end{split}
\end{equation}
where $r$ is the radius within a dark matter halo and $M_0 = 10^{10} M_\odot {\rm h}^{-1}$. $A_{\rm \hi}$ is a free parameter that controls the amplitude of the \hi-halo mass relation, $\beta$ determines the slope of the relation, $\gamma$ determines the slope of the \hi\ mass profile inside the halos. Together with the velocity dispersion $\sigma_v$ and the shot noise $P_{\rm SN}^0$, they form the 5-parameter model which we use for forecasts. For simplicity, we present the forecasts only for the $z\sim 1$ bin.
The fiducial values adopted are listed in \autoref{tab:pars}.

\begin{table}[]
    \centering
    \begin{tabular}{c|c|c|c|c|c|c}
    \hline
       & ${\rm log}_{10}[A_{\rm HI}/M_\odot]$ & $\beta$ & $\gamma$ & $\sigma_v\,$[km/s] & $P_{\rm SN}^0\,[{\rm Mpc^{3}}]$ & $\Omega_{\rm \hi} \times 10^4$  \\\hline
    Prior & (7,10) & (0,4) & (1,3) & (0,250) & (1,500) & (1,100) \\
    Fiducial & 8.600 & 0.700 & 1.50 & 50.00 & 120.00 & 4.95 \\
    Forecast & $8.599^{+0.004}_{-0.007}$ & $0.697^{+0.033}_{-0.030}$ & $1.48^{+0.28}_{-0.26}$ & $49.93^{+5.85}_{5.77}$ & $118.10^{+15.22}_{-13.58}$ & $4.96^{+0.12}_{-0.14}$ 
    \end{tabular}
    \caption{The forecasts for halo model constraints using SKA-Mid interferometric \hi\ intensity mapping at $z\sim1$. The first row lists the model parameters and the derived quantity $\Omega_{\rm \hi}$.
    {The second row lists the boundary of the flat priors for the parameters.}
    The third row lists the fiducial values of the parameters.
    The fourth row lists the forecasted constraints with the {[16\%, 50\%, 84\%]} confidence interval.}
    \label{tab:pars}
\end{table}

\begin{figure}[h]
    \centering
	\includegraphics[width=0.9\columnwidth]{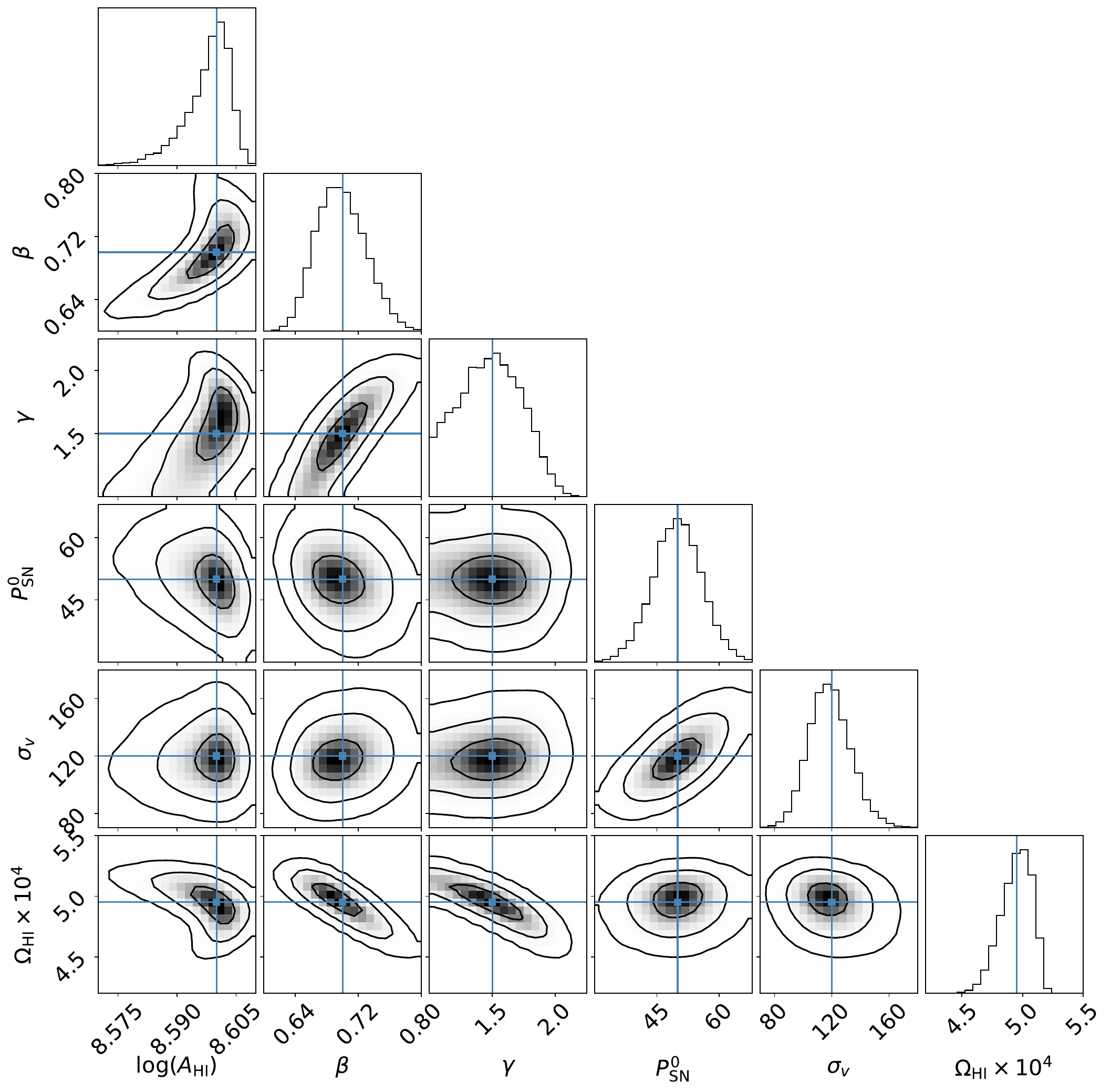}
    \caption{Forecasts on the posterior distribution of the halo model parameters for the interferometric \hi\ intensity mapping measurements for SKA-Mid AA4 configuration at $z\sim 1$. The contours show the $1\sigma$, $2 \sigma$ and $3\sigma$ confidence level, respectively. The solid lines denote the fiducial values listed in \autoref{tab:pars}.}
    \label{fig:cornerhalo}
\end{figure}

Using the fiducial model and the measurement error presented in \secref{sec:forecast}, we perform Bayesian inference on the model parameters, assuming a diagonal Gaussian likelihood.
We use \textsc{nautilus} \citep{nautilus} to perform importance nested sampling.
{We adopt wide, flat priors for all model parameters listed in \autoref{tab:pars}. The $\gamma$ parameter has a prior of (1,3) as values outside this region gives unphysical divergence.
We choose the number of live points to be 2000 and consider the sampling to have converged if the evidence in the live set is less than 1\% of the total evidence (see Eq. 8 of \citealt{nautilus}).}
The resulting $1\sigma$ confidence interval is listed in \autoref{tab:pars}, and the posterior presented in \autoref{fig:cornerhalo}. As shown, the halo model parameters can be precisely constrained, resulting in unprecedented measurements on the relation between the dark matter halos and their \hi\ content.
As we demonstrate in \autoref{fig:cornerhalo}, despite parameter degeneracies, the high signal-to-noise measurements of SKA-Mid will be able to break the parameter degeneracy and provide unbiased estimation of the model parameters.
In particular, extremely accurate measurements of the \hi-halo mass relation will be probed, providing important insights into hydrodynamical simulations of cosmic structure.
Moreover, the halo model constraints can in turn produce an accurate measurement of the \hi\ density $\Omega_{\rm \hi}= 4.96^{+0.12}_{-0.14} \times 10^{-4}$, extending \hi\ density measurements using emission line to $z\sim 1.0$ and beyond.

\subsection{HIMF}
The halo model constraints from the measured \hi\ power spectrum can be turned into constraints on the HI mass function (HIMF), as suggested in \cite{2021MNRAS.502.5259C}. The HIMF is typically parameterised as \citep{Schechter_1976}:

\begin{equation}
    \phi_{\rm \hi}(M_{\rm \hi})\equiv \frac{{\rm d}n_{\rm \hi}}{{\rm dlog}M_{\rm \hi}} = {\rm ln}10\,\phi_* \big(\frac{M_{\rm \hi}}{M_*}\big)^{\alpha +1}e^{-\frac{M_{\rm \hi}}{M_*}},
\label{eq:himf}
\end{equation}
where $[\phi_*,M_*,\alpha]$ are the parameters that control the amplitude, knee mass and the slope of the HIMF respectively.

The HIMF can then be used to calculate the \hi\ density and the real-space shot noise \citep{2021MNRAS.502.5259C}:
\begin{equation}
    \Omega_{\rm \hi} = \int {\rm dlog}M_{\rm \hi}\,\phi_{\rm \hi}(M_{\rm \hi}) M_{\rm \hi}/\rho_c
\label{eq:omegahihimf}
\end{equation}
\begin{equation}
    P_{\rm SN}^0 = (C_{\rm \hi}\rho_c \Omega_{\rm \hi})^2 \frac{\int {\rm dlog}M_{\rm \hi}\,\phi_{\rm \hi}(M_{\rm \hi})M_{\rm \hi}^2}{\big(\int {\rm dlog}M_{\rm \hi}\,\phi_{\rm \hi}(M_{\rm \hi})M_{\rm \hi}\big)^2}
\label{eq:psn}
\end{equation}

\autoref{eq:himf} contains three HIMF parameters. In principle, the \hi\ power spectrum provides the measurements of the overall \hi\ density as well as the shot noise. Combining \hi\ power spectrum with complimentary measurements, such as stacking, can be used to fully constrain the HIMF. In this chapter, we focus on using only the \hi\ power spectrum. In this case, the HIMF parameters will not be fully constrained, due to the degeneracy of HIMF parameters in $\Omega_{\rm \hi}$ and $P_{\rm SN}^0 $. However, the parameter posterior still gives informative constraints on the shape of the HIMF, especially on the ``knee mass'' of the HIMF.

\begin{figure}[h]
    \centering
	\includegraphics[width=0.49\columnwidth]{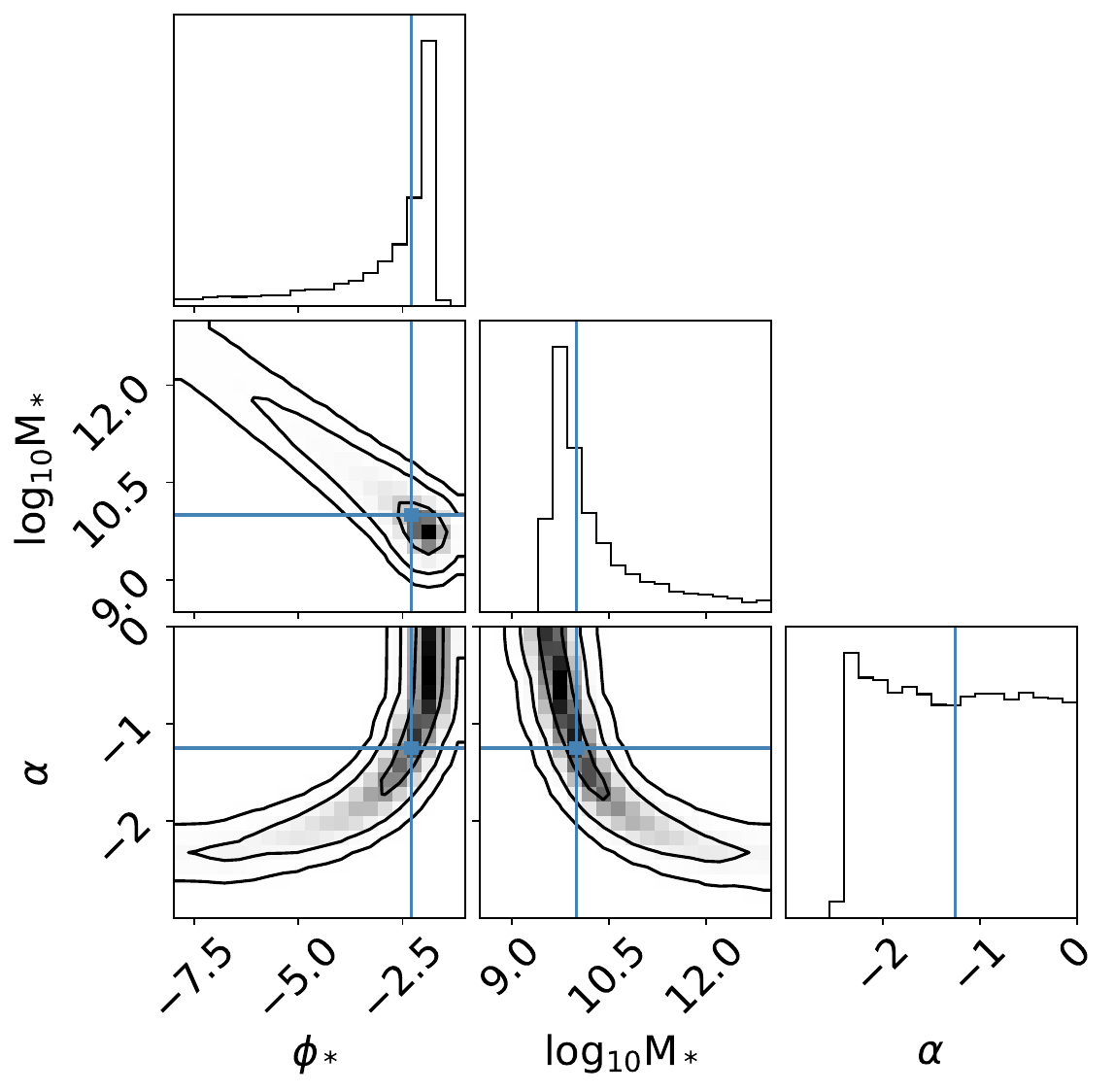}
	\includegraphics[width=0.49\columnwidth]{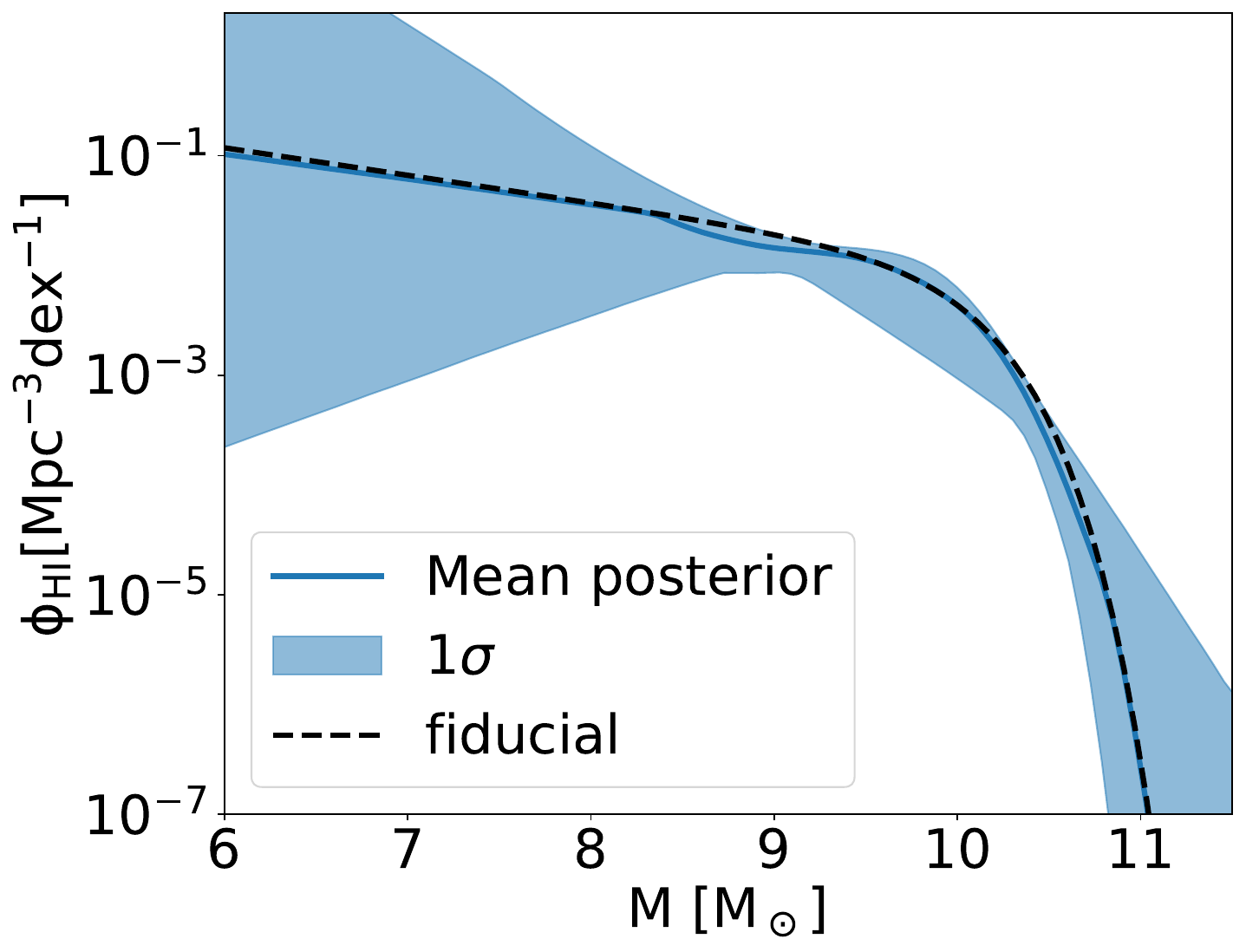}
    \caption{Left panel: Forecasts on the posterior distribution of the HIMF parameters for SKA-Mid AA4 configuration at $z\sim 1$. The contours show the $1\sigma$, $2 \sigma$ and $3\sigma$ confidence level, respectively.
    Right panel: Forecast on the posterior distribution of the measured HIMF. The black dashed line denotes the fiducial HIMF. The blue solid line denotes the mean of the posterior.
    The shaded region denotes the 1$\sigma$ confidence interval of the measurement.
    }
    \label{fig:cornerhimf}
\end{figure}

We repeat the same Bayesian inference routine in \secref{subsec:halomodel}, and substitute the shot noise free parameter with \autoref{eq:psn}.
$M_*$ and $\alpha$ parameters are sampled with wide flat priors {of $\log [M_*/ M_\odot]\in (6,13)$ and $\alpha \in (-5,0)$}, whereas the $\phi_*$ is calculated in each sample point based on $\Omega_{\rm \hi}$ according to \autoref{eq:omegahi} and \autoref{eq:omegahihimf}.
The results are shown in \autoref{fig:cornerhimf}.
The HIMF parameter constraints are significantly degenerate due to the lack of information on the number density of \hi\ sources from intensity mapping.
The slope of the HIMF is not sensitive to intensity mapping measurements and is {largely unconstrained, except for a lower limit of $\alpha \gtrsim -2.5$}.
The amplitude and the knee mass of the HIMF, on the other hand, can be measured.
The reconstructed HIMF function is shown in the right panel of \autoref{fig:cornerhimf}.
As shown, the posterior can reconstruct the overall shape of the HIMF, especially in the mass range around the knee mass.
The results demonstrate that \hi\ intensity mapping with SKA-Mid can be used to constrain HIMF, pushing the measurements beyond previous limits $z\lesssim 0.5$.

\subsection{Progress in modelling non-linear power spectrum}

The anticipated science outputs shown above demonstrate the power of \hi\ intensity mapping, and the importance of modelling the \hi\ power spectrum at non-linear scales.
In particular, the modelling of the \hi\ shot noise is important, as the shot noise dominates the clustering amplitude in small scales, and enables constraints on the HIMF.

In a low-redshift universe ($z\leq 1.0$), most of the atomic hydrogen (\hi) gas resides in the cold dense region of galaxies, where it can be self-shielded from ultraviolet photons \citep[see e.g.,][]{Krumholz2009, Diemer2018, Villaescusa-Navarro2018}. Ideally, particle information is needed to obtain the non-linear ($k\gtrsim 1\ {\rm Mpc^{-1}}$) power spectrum given the scales. However, there might not be much information contained in the power spectrum with very small scales (say, $k\gtrsim 10\ {\rm Mpc^{-1}}$). For the scale of our concern ($1\ {\rm Mpc^{-1}}\lesssim k\lesssim 10\ {\rm Mpc^{-1}}$), several factors might play a role, including the diameter of \hi\ discs, stretching length of \hi\ velocity dispersion in redshift space and the distances between galaxies. When modelling the \hi\ field for some specific surveys, it is strongly recommended to compare the scale of different factors with the resolution or the scale of your concern along and perpendicular to the line-of-sight direction, and then to determine which factors to be modelled.

Take the detection by \cite{Paul2023} as an example, \cite{Li-ZX2024} compares different factors qualitatively by assuming that the \hi\ statistics would not change dramatically as $z$ changes from $0.0$ to $1.0$. It is found that the sizes of \hi\ discs have subtle effects, while other factors are more important. Although the \hi\ IM surveys performed by the MeerKAT interferometers do not identify individual sources as the extragalactic surveys do, these sources can still be resolved by the surveys and leave some imprints on the measured nonlinear \hi\ power spectrum. Specifically, sub-structures of \hi\ insides halos alter the shot noise term of the power spectrum, and galaxy peculiar velocities exacerbate the Finger of God effect. Hence, halo models without substructure modelling are not ideal for predictions on nonlinear scales. Velocity dispersions due to the \hi\ gas random motions and circular rotations suppress the non-linear power spectrum, especially the shot noise term. \cite{Li-ZX2024} further finds that the \hi\ power spectrum is relatively insensitive to the profile shape of the \hi\ emission line at these scales, while showing a strong correlation with the profile width. By assuming a Gaussian shape for each emission line profile, the shot noise term can be modelled as Equation \ref{eq: sn term}.

\begin{equation}
\label{eq: sn term}
    P_{\rm SN}(k,\mu)=\dfrac{V}{M_t^2}\cdot\sum_i M_i^2\cdot e^{-\sigma_i^2\mu^2k^2}\ \ \ 
    ,\ \ \ \mu=k_{z}/k
\end{equation}
where $V$ is the volume of the simulated box; $M_{\rm t}$ is the total mass of all \hi\ sources; $M_i$ is the \hi\ mass of each galactic \hi\ source; $\sigma_i$ is the $\sigma$ parameter in the Gaussian profiles. $\sigma_i$ is used to describe the profile width and is thus determined by the Full Width Half Maximum (FWHM) $w_{\rm 50}$ obtained in extragalactic \hi\ surveys as Equation \ref{eq: sigma_i}.

\begin{equation}
\label{eq: sigma_i}
    \sigma_i = \dfrac{l_{w50}}{2\sqrt{2\ln{2}}},
\end{equation}
where the $l_{w50}$ is the stretched length due to the profile width in the redshift space.

If we have a 2-dimensional distribution $\phi(M, w_{\rm 50})$, Equation \ref{eq: sn term} can also be written as Equation \ref{eq: sn term v2}. It is apparent that shot noise term have strong correlation with the 2D \hi\ distribution $\phi(M, w_{\rm 50})$ obtained in extragalactic surveys, and shot noise term in the power spectrum can thus provide constraint on the \hi\ velocity function and also \hi\ mass function.

\begin{equation}
\label{eq: sn term v2}
    P_{\rm SN}(k,\mu)=V\cdot\dfrac{\int\int \phi(M, w_{\rm 50})M^2\cdot e^{-\sigma(w_{\rm 50})^2\mu^2k^2}\ {\rm d}M {\rm d}w_{\rm 50}}{\int\int \phi(M, w_{\rm 50})M\ {\rm d}M {\rm d}w_{\rm 50}}\ \ \ 
    ,\ \ \ \mu=k_{z}/k.
\end{equation}

Additionally, although it is easily overlooked, the observational settings, like horizon limit and frequency resolution, can have a relatively large impact on the results \citep[see Figure 11 in ][]{Li-ZX2024}. Different surveys might utilise different observational settings, so forecasting based on specific settings of the different surveys is always essential.

\subsection{Cross-correlation with galaxy surveys}
\label{sec:cross}

The 21-cm signal traces the neutral gas distribution over a wide redshift range, probing cosmological volumes of the order of a few tenths of $\mathrm{Gpc^3}$ around $z \sim 1$. The inherently higher resolution of radio data in both spatial and radial directions (compared to optical spectroscopic surveys) would enable efficient mitigation of instrumental and astrophysical effects originating from separate sources in optical and radio data through cross-correlation measurements. While such investigations are routinely conducted with single dish maps (see \cite{Cunnington01.2026.SKA} for a discussion of progress with MeerKAT), direct cross-correlation between interferometric intensity maps and other large-scale structure tracers is yet to be achieved, {though there have been positive results using stacking out to redshift $\sim$1 (for example \cite{chime_galaxy}). In this section, we discuss the prospects of cross-correlation without using stacking}. 

\begin{figure}
    \centering
	\includegraphics[width=0.8\columnwidth]{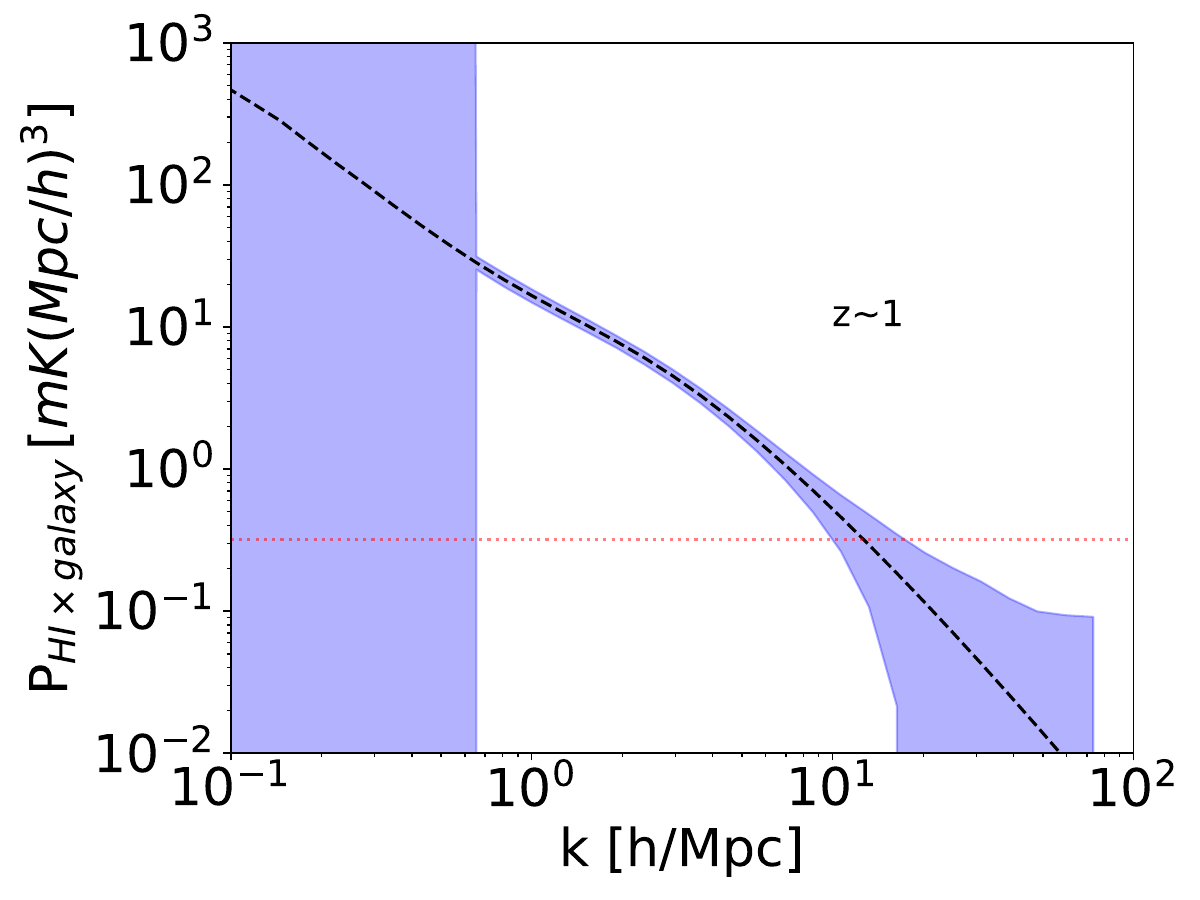}
    \caption{Forecasts on the uncertainty on the \hi-galaxy cross-correlation power spectrum using galaxy number density for Euclid survey (10$^{-4}$) and a 1000\,hour survey with the Band-1 of SKA-Mid using {AA4} configuration. The grey dotted line shows the cross-power amplitude, and the red dashed line shows the shot-noise levels.}
    \label{fig:forecast_cross}
\end{figure}

As discussed earlier, the prospects of intensity mapping are very promising with the SKA-Mid owing to its compact baselines design and excellent noise characteristics. Using these datasets, we can realistically forecast the prospects of cross-correlation with galaxy redshift surveys. Several of the surveys with SKA-Mid will overlap in sky coverage with Stage IV optical surveys such as Euclid and LSST. Thus, a realistic cross-correlation measurement can be achieved with a survey covering only a few hundred square degrees. \autoref{fig:forecast_cross} shows the forecasts on the uncertainties of the cross-power measurements for such a scenario. It assumes an SKAO ``medium-deep'' type radio survey covering 100 square degrees with 1000\,hours observing time and an overlapping survey with a volume density 10$^{-4}$ for galaxy number counts. The forecast assumes a redshift range $0.5 \lesssim z\lesssim1.5$, accessible with the Band-1 of SKA-Mid. As seen in \autoref{fig:forecast_cross}, the uncertainties achieved under these realistic survey specifications are quite low between scales $\sim$0.5 to 10 $hMpc^{-1}$. Beyond $\sim$10 $hMpc^{-1}$, the cross-power starts being shot noise dominated, hence, enhanced observing times are of little help. However, the additive shot noise in the cross-power spectrum is also a useful probe of gas astrophysics. {It is to be noted that the optical galaxy number density is representative of the Euclid survey \citep{2020A&A...642A.191E}, and the expected signal-to-noise would vary depending on the specifications of the galaxy survey used. A lower sensitivity or shallower survey may reduce the galaxy number counts or enhance the shot noise levels or both, which affects the SNR. Thus, such a survey with higher uncertainties on the cross-power spectrum would lead to looser constraints on the subsequent parameters estimated. However, these forecasts are still valid under the realistic condition that the patch of the sky with radio data is also covered by Euclid.} 

{Cross-correlation measurement can be used as a complementary tool to auto-correlation for not only parameter estimation but also diagnostics. While cross-correlation alone would not independently measure the full clustering amplitude, different tracers affected by uncorrelated systematic biases can help us quantify them (for example, any bias due to foreground residuals in {\hi} auto-power spectrum would be largely cancelled in the cross-power). Sections 4.1 to 4.3 discuss the constraints on global {\hi} properties using auto-correlation data. However, using scales accessible with interferometers, cross-correlations with optical galaxy catalogues can also be used to determine the average {\hi} content in galaxies. Galaxy catalogues binned using different selection criteria, such as stellar mass and colour, can be used to probe {\hi} scaling relations. Further, \cite{2017MNRAS.470.3220W} showed that the additive shot noise term in the cross-power spectrum scales with the average {\hi} brightness temperature of the optically selected galaxies. Thus, the power spectrum obtained from cross-correlation between {\hi} intensity maps and optical galaxies on small scales presents a unique opportunity for studying galaxy evolution independently.} 

Further, cross-correlation of \hi\ intensity maps with Ly-$\alpha$ data from spectroscopic surveys will help remove some of the degeneracies. One of the important contaminants in Ly-$\alpha$ data inhibiting cosmological interpretations is the ``high-column density systems'' (HCDs), dense clouds of \hi\ associated with the circumgalactic medium. In contrast, the Lyman-alpha forest cosmological analysis focuses on the dilute intergalactic medium. The \hi\ intensity-mapping signal is expected to be dominated by \hi\ gas in galaxies, and hence highly correlated with the HCDs. It should be possible to estimate the impact of HCDs on the cosmological constraints from the Ly-$\alpha$ forest analysis by cross-correlating the optical spectroscopic data with 21-cm intensity maps. Overall, the SKAO presents unprecedented opportunities to leverage cross-correlations between different large-scale structure probes, at least to redshift $\sim$1.


\section{Conclusion}
This chapter describes the progress in post-EoR \hi\ intensity mapping with radio interferometers. At the redshift range in question, while single dish surveys are sensitive to the large angular scales, interferometers are sensitive to the smaller angular separations. Hence, observation and modelling in these scales probe the neutral gas astrophysics as well as the relationship between the gas and the host haloes. There has been significant progress in recent years in both observations and theoretical modelling. 

The main highlights of the milestones from observations and systematic modelling include:

\begin{enumerate}
    \item A tentative \hi\ power spectrum detection reported using $\sim$100\,hours deep observations of MeerKAT telescope. Observations targeted redshifts centred around $z\sim0.32$ and $z\sim0.44$. The calibration and power spectrum pipelines developed enabled robust delay-spectrum estimation directly from visibilities.
    \item An upper limit on the \hi\ power spectrum was placed using wide-area observations of the MeerKAT MIGHTEE survey. Using the incoherent averaging technique to average data from multiple pointings in the Fourier domain, this work used a redshift bin centred at $z\sim0.44$ and demonstrated the feasibility of using wide area surveys. 
    \item An investigation on testing the performance of intensity mapping, using \hi\ sources directly detected from MIGHTEE observations, is demonstrated for $z<0.1$. It allows us to test the robustness of the methodology developed to investigate how much \hi\ either the intensity mapping signal or galaxy detections miss. The work demonstrates using intensity mapping and galaxy detections that, in the limit of the galaxy mesh resolution, the power spectrum measurements from interferometric intensity mapping data agree well with the \hi\ galaxy catalogue, but become unreliable at scales smaller than the resolution.
    \item There have also been developments of several systematic isolation methods, such as using baseline and $uv$-domain flagging, using the 2D Fourier plane \& development of MASS, a simulator for estimating visibilities and systematics from MeerKAT.
\end{enumerate}

Parallelly, there have also been developments in modelling the non-linear power spectrum and constraints on the halo model of cosmic \hi. Forecasts for SKA data using a standard survey of 100 trackings of different pointings, each with 10\,hrs of integration time, show promising results for recovery of the \hi\ power spectrum and subsequent estimation of halo model parameters and the \hi\ mass function.
Using the halo model formalism, constraints of the \hi\ density $\Omega_{\rm \hi}$ can be achieved at per-cent level accuracy at $z\sim 1.0$.
Measurements of the \hi\ density and the \hi\ shot noise can be used to measure the HIMF at $z\sim 1.0$, beyond limits of conventional \hi\ stacking techniques.
Interferometric intensity mapping with SKA-Mid will be able to open a new window into probing the cosmic \hi\ content and the galaxy evolution at $1.0\lesssim z \lesssim 3.0$.

\bibliographystyle{abbrvnat-maxbibnames4}
\bibliography{chapter} 

\end{document}